\documentclass[sigconf,natbib=true]{acmart}
\usepackage[ruled,linesnumbered]{algorithm2e}
\usepackage{multirow}
\usepackage{makecell}
\usepackage{enumitem}
\setitemize[1]{noitemsep,partopsep=0pt,parsep=0pt,topsep=0pt, leftmargin=10pt,
rightmargin=0pt}
\AtBeginDocument{%
  \providecommand\BibTeX{{%
    \normalfont B\kern-0.5em{\scshape i\kern-0.25em b}\kern-0.8em\TeX}}}

\setcopyright{acmcopyright}
\copyrightyear{2018}
\acmYear{2018}
\acmDOI{10.1145/1122445.1122456}

\acmConference[Woodstock '18]{Woodstock '18: ACM Symposium on Neural
  Gaze Detection}{June 03--05, 2018}{Woodstock, NY}
\acmBooktitle{Proceedings of the 47th International ACM SIGIR Conference on Research and Development in Information Retrieval (SIGIR),
  July 14--18, 2024, Washington D.C., USA}
\acmPrice{15.00}
\acmISBN{978-1-4503-XXXX-X/18/06}

\begin{document}

%%
%% The "title" command has an optional parameter,
%% allowing the author to define a "short title" to be used in page headers.
\title{Scenario-Adaptive Fine-Grained Personalization Network: Tailoring User Behavior Representation to the Scenario Context}

%%
%% The "author" command and its associated commands are used to define
%% the authors and their affiliations.
%% Of note is the shared affiliation of the first two authors, and the
%% "authornote" and "authornotemark" commands
%% used to denote shared contribution to the research.

\author{Moyu Zhang}
\affiliation{%
  \institution{Lazada Group}
  \city{Beijing}
  \state{Beijing}
  \country{China}
}
\email{zhangmoyu@bupt.cn}

\author{Yongxiang Tang}
 % \authornotemark[1]
\affiliation{%
  \institution{Unaffiliated}
    \city{Beijing}
  \state{Beijing}
  \country{China}
}
\email{tangyongxiang94@gmail.com}

\author{Jinxin Hu}
 % \authornotemark[1]
 \authornote{Corresponding Author}
\affiliation{%
  \institution{Lazada Group}
    \city{Beijing}
  \state{Beijing}
  \country{China}
}
\email{jinxin.hjx@lazada.com}

\author{Yu Zhang}
 % \authornotemark[1]
\affiliation{%
  \institution{Lazada Group}
    \city{Beijing}
  \state{Beijing}
  \country{China}
}
\email{daoji@lazada.com}

%% The abstract is a short summary of the work to be presented in the article.
\begin{abstract}
As e-commerce has evolved, modern large-scale commercial platforms now accommodate various scenarios to cater to the diverse shopping preferences of users. To conserve resources, current multi-scenario methods often utilize a unified framework to deliver personalized recommendations across various scenarios. Given the overlap of users and items in multiple scenarios on commercial platforms, current methods typically employ shared bottom representations, capturing similarities and differences between scenarios through adaptive adjustments. However, existing methods often adjust representations adaptively only after aggregating user behavior sequences. This coarse-grained approach to re-weighting the entire user sequence hampers the model's ability to accurately model the user interest migration across different scenarios. To enhance the model's capacity to capture user interests from historical behavior sequences in each scenario, we develop a ranking framework named the \textbf{S}cenario-Adaptive \textbf{F}ine-Grained \textbf{P}ersonalization \textbf{Net}work (SFPNet), which designs a kind of fine-grained method for multi-scenario personalized recommendations. Specifically, SFPNet comprises a series of blocks named as \emph{Scenario-Tailoring Block}, stacked sequentially. Each block initially deploys a parameter personalization unit to integrate scenario information at a coarse-grained level by redefining fundamental features. Subsequently, we consolidate scenario-adaptively adjusted feature representations to serve as context information. By employing residual connection, we incorporate this context into the representation of each historical behavior, allowing for context-aware fine-grained customization of the behavior representations at the scenario-level, which in turn supports scenario-aware user interest modeling. Ultimately, the effectiveness of our proposed method is strongly substantiated by extensive experiments and online A/B testing. 
\end{abstract}

\keywords{Multi-Scenario Recommendation, Scenario-Specific Behavior Representation, Recommender System}

\maketitle

\begin{figure}[t]
  \centering
  \includegraphics[width=\linewidth]{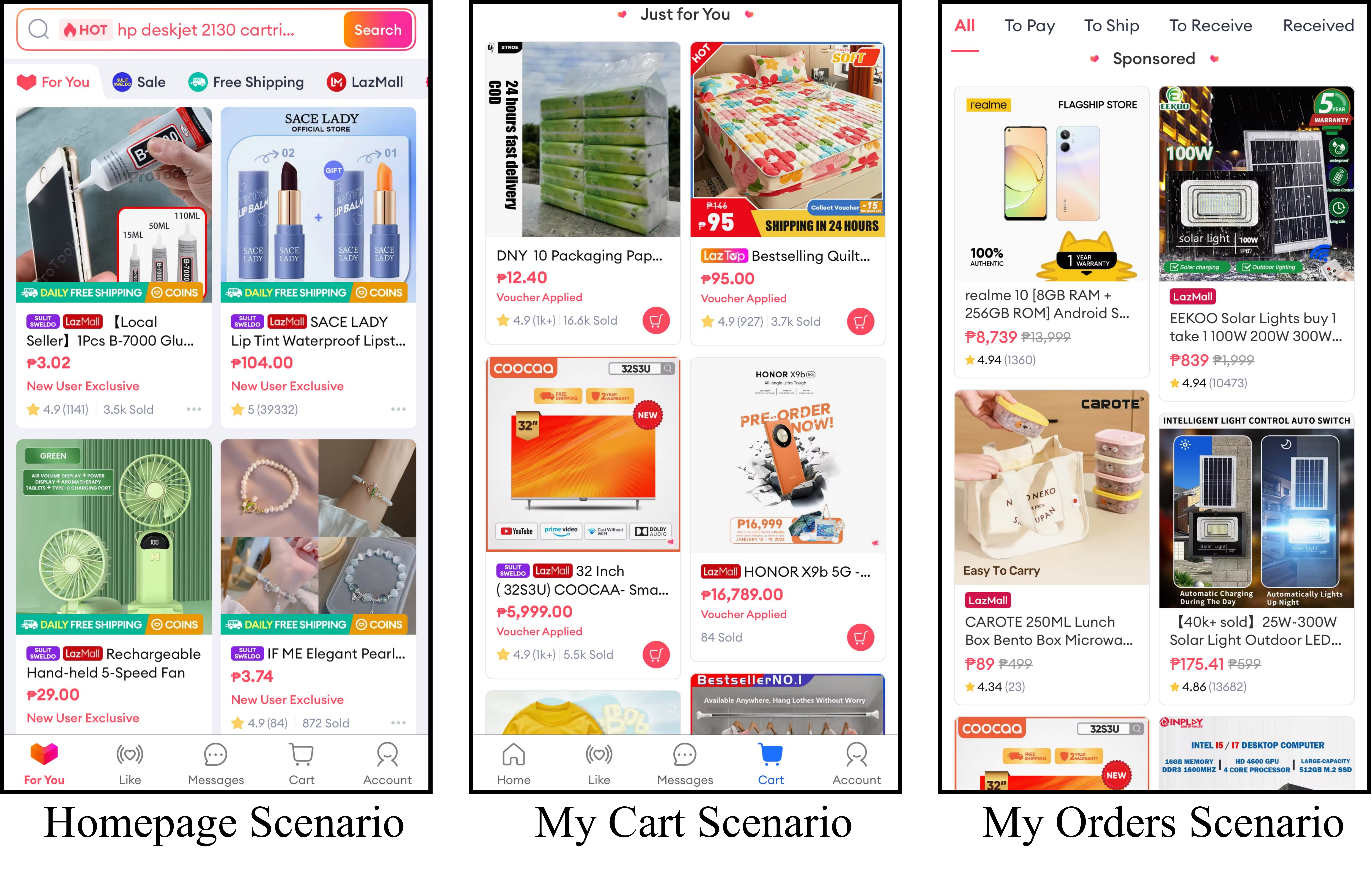}
  \caption{An illustrative example of multi-scenarios for the online e-commerce platform Lazada.}
  \label{example_scene}
\end{figure} 

\section{Introduction}
In recent years, with the rise of electronic shopping as a dominant trend, a multitude of e-commerce platforms have emerged, including Amazon, Alibaba's Taobao, Lazada, TikTok, and others \cite{back1, back2, back3, back4, causalint, adl}. To cater to the growing diversity of user shopping preferences, contemporary commercial platforms must offer personalized services tailored to a variety of shopping contexts.  For instance, the online e-commerce platform depicted in Figure \ref{example_scene} offers multiple item recommendation scenarios. The "Homepage" scenario primarily suggests a broad range of products potentially appealing to the user, based on their past behavior, whereas the "My Cart" and "My Orders" scenarios depend more on the user's historical purchase records to generate recommendations. Compared to the "Homepage" scenario, user interests in the "My Cart" and "My Orders" scenarios become more specific. Consequently, these multi-scenario platforms present a challenge to existing recommendation models: how to devise personalized recommendations that consider the distinct characteristics of each scenario?

Traditional methods typically employ scenario-specific data to train distinct models for each scenario, enhancing scenario differentiation for online services. However, this isolated model training approach is resource-intensive, often inefficient, and fails to capture inter-scenario correlations, leading to suboptimal performance in data-sparse scenarios. In practice, most current e-commerce platforms experience significant user and item overlap across various scenarios. Consequently, despite the differences in user and item performance across scenarios, substantial correlations persist. Leveraging the correlations across scenarios, we can incorporate data from other scenarios into the training process, alleviating the challenge of data sparsity in individual scenarios. To model these inter-scenario correlations, mainstream multi-scenario approaches often employ a unified recommendation modeling framework, broadly categorized into two types: 1) Scenario-specific network structures, influenced by multi-task learning (MTL) \cite{mtl1, mtl2, mtl3}. Each scenario is treated as a distinct task, with a common network capturing inter-scenario correlations and separate networks modeling the unique aspects of each scenario. 2) Scenario-adaptive parameter network structures \cite{pepnet, adasparse}, which differ from MTL by acknowledging feature space variations. These methods apply scenario data directly to the core embedding and prediction layers, enabling dynamic adjustments of the feature space and prediction strategy in response to scenario shifts. Compared to MTL methods that require learning distinct network structures for new scenario, the adaptive parameter approach simply incorporates new scenario-related features, enhancing its flexibility to recognize new scenarios.

\begin{figure}[t]
  \centering
  \includegraphics[width=\linewidth]{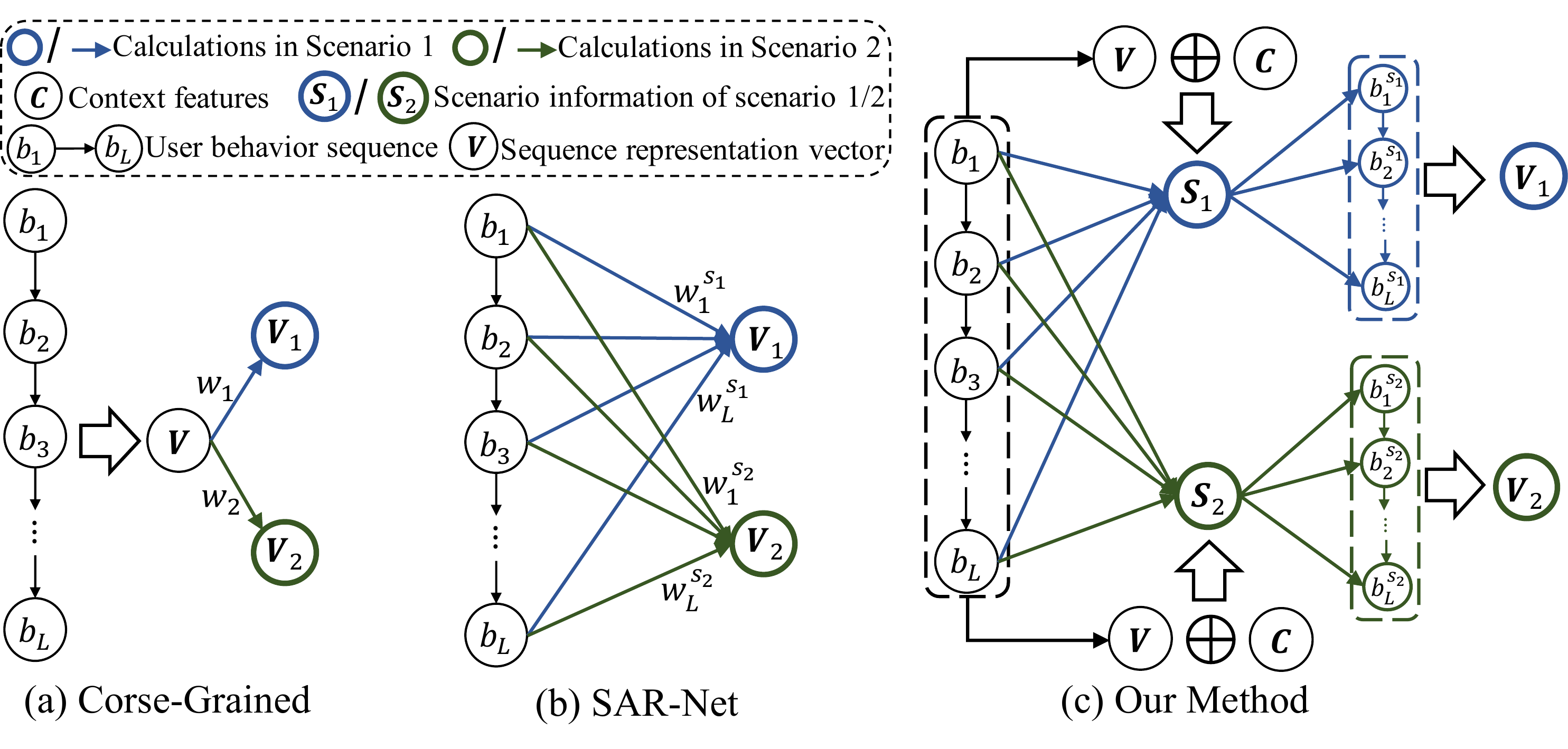}
  \caption{Examples of previous methods and ours. In (a), $\boldsymbol{V}_1=w_1\boldsymbol{V}$ and $\boldsymbol{V}_2=w_2 \boldsymbol{V}$, where $\boldsymbol{V} = f(\boldsymbol{b}_1, ..., \boldsymbol{b}_L)$. While in (b), $\boldsymbol{V}_1=f(w_1^{s_1}\boldsymbol{b}_1, ..., w_L^{s_1}\boldsymbol{b}_L)$ and $\boldsymbol{V}_2= f(w_1^{s_2}\boldsymbol{b}_1, ..., w_L^{s_2}\boldsymbol{b}_L)$. In our method, $\boldsymbol{V}_1=f(\boldsymbol{b}_1^{s_1}, ..., \boldsymbol{b}_L^{s_1})$, where $\boldsymbol{b}_1^{s_1}$ is obtained by $(\boldsymbol{V}, \boldsymbol{C}, \boldsymbol{S}_1, \boldsymbol{b}_1)$.}
  \label{example_seq}
\end{figure}

However, current scenario-adaptive parameter structures face a common issue: they generally transfer information between scenarios coarsely, treating the user's entire historical behavior sequence as a single feature for adjustments, which fails to consider the migration of user interests with the scenario changes. As depicted in Figure \ref{example_seq}(a), the weight adjustment method uniformly applies the same weight to each behavior in the user sequence, rather than tailoring scenario-specific information to individual behaviors, thereby impairing the model's capacity to accurately track user interest shifts across scenarios. In fact, modeling user interests is critical in recommendation systems, particularly as users do not always reveal their intentions in non-search scenarios, making it essential to accurately track the migration of user interests for effective multi-scenario recommendations. For instance, on the "Homepage" scenario, where user interests tend to be broad, considering each click behavior within their historical behaviors is necessary to provide diverse recommendations. Conversely, in scenarios guided by specific interests, users may exhibit more defined preferences, necessitating a focus on corresponding behaviors within their sequence. Therefore, to model user interests across varied scenarios more effectively, a fine-grained analysis of the relationships between the user's historical behaviors and scenarios is essential. Although the SAR-Net model \cite{sar} ever attempted to incorporate scenario information into sequence modeling using a target-attention mechanism \cite{din}, as shown in Figure \ref{example_seq}(b), this approach of weighted summation does little to alter the actual representation of the user sequence. Its ability to represent the complex interplay between real-life scenarios and user behaviors is markedly limited. Consequently, adaptively and reasonably adjusting the representations of user historical behavior to match various scenarios remains an urgent challenge in contemporary multi-scenario modeling.

To tackle the above challenge of multi-scenario modeling, we propose a new framework, the \textbf{S}cenario-Adaptive \textbf{F}ine-Grained \textbf{P}ersonalization \textbf{Net}work (SFPNet), designed to finely capture user interest changes across different scenarios and enhance recommendation accuracy in a multi-scenario context. SFPNet features a deep network architecture built from a series of stacked blocks named as \textbf{\emph{Scenario-Tailoring Block}}. Each block consists of two critical modules: the Scenario-Adaptive Module and the Residual-Tailoring Module. The former module adjusts foundational features to serve as instance-specific contextual information, facilitating scenario awareness, while the latter module uses this contextual information to tailor scenario representations for each feature.

Specifically, as identified in numerous studies \cite{pepnet, maria}, the foundational feature representation vector is frequently deemed the most critical bottleneck in recommendation models. Therefore, SFPNet utilizes a \emph{Scenario-Adaptive Module (SAM)} that dynamically scales foundational features. This is achieved by inputting scenario prior features into a scenario-specific gate, followed by an element-wise multiplication with the original feature representations. These scaled representations embody scenario-aware contextual information, signifying the input instance's presence in the current scenario. Notably, in the SAM, calculations are performed after compressing the sequence into a multi-dimensional vector, which aims to enhance the perception of the sequence's contextual information for subsequent fine-grained behavioral modeling. Upon acquiring the instance context information, we devised a \emph{Residual-Tailoring Module (RTM)} in response to concerns that the vector-level re-weighting summation to model the shifts in user interests as reflected by user sequences across different scenarios, as utilized by SAR-Net \cite{sar}, may inadequately capture the intricate relationship between the scenario and user interests. To facilitate scenario-level customization for each behavior in the sequence, we employ residual connections \cite{res} that depart from traditional feature-scenario interactions. Instead, RTM integrates the contextual information of the input instance and leverage a neural network to dynamically generate high-order interactions, subsequently tailoring the behavior representation of the sequence via residual computations. The integration of the above two modules allows SFPNet to incorporate scenario-related information into the feature representation vectors with high granularity, markedly enhancing the model's proficiency in discerning commonalities and distinctions across scenarios.

The contributions of our paper can be summarized as follows:
\begin{itemize}
\item To our knowledge, SFPNet is the first multi-scenario work to offer fine-grained scenario-specific customization for individual user behaviors, thereby advancing the model's capacity to track shifts in user interests across various scenarios.
\item SFPNet features a novel block comprising two key modules, which capitalizes on the context and scenario-specific information of each input instance to tailor the representation vectors of user behaviors in the sequence and associated context features of the input instance. By stacking the block, SFPNet constructs a deep network that progressively bolsters its ability to model the intricate interplay between scenarios and input instances.
\item Evaluations using offline datasets and online A/B testing have demonstrated the superiority of the proposed SFPNet method over existing state-of-the-art multi-scenario approaches.
\end{itemize}  
 
\section{Related Work}
With the evolution of e-commerce platforms and growing user engagement, a single scenario is insufficient to satisfy the expanding needs of users \cite{m5, samd, hc2}. Consequently, modern online e-commerce platforms are required to cater to multiple scenarios simultaneously. Initially, recommendation systems typically trained a distinct model for each scenario to cater to its specific demands. However, with the proliferation of platform scenarios, the drawbacks of this independent training approach have become increasingly evident. This method is not only resource-intensive but also fails to leverage the correlations between scenarios, leading to suboptimal model performance. Recent advances in deep learning and the industry's emphasis on replicability—avoiding the integration of new elements with each additional scenario—have led to a preference for a unified model structure, employing a singular ranking framework for scenario-specific recommendations. Single model structures are primarily categorized into two approaches: 1) Scenario-specific network structures, and 2) Scenario-adaptive parameter network structures. Though both strive to train a unified model capable of serving multiple scenarios, their underlying philosophies differ. 

Specifically, scenario-specific network structures draw inspiration from multi-task learning (MTL) principles. These approaches seek to harness the similarities between multi-task and multi-scenario applications by considering each scenario as an independent task. Mixture-of-Experts (MoE) \cite{moe} is proposed to significantly increase model capacity and capability and is widely used in the multi-scenario area. MMoE \cite{mtl3} characterizes the task correlation and learns the function of specific tasks based on shared bottom feature representation. HMoE \cite{hmoe} takes advantage of MMoE to implicitly identify distinctions and commonalities between tasks, and improves the performance with a stacked model learning task relationships in the label space explicitly. Later, PLE \cite{ple} shares experts in the share layer and refines tasks uniquely to effectively alleviate the noise caused by other scenarios and improves the effectiveness of feature extraction. M2M \cite{m2m} utilizes expert networks to solve multi-scenario and multi-task problems based on a meta network to express the scenario information explicitly. AESM$^2$ \cite{aesm2} proposes a novel expert network structure with automatic selection of fine granularity by calculating the KL divergence to select the most suitable sharing and exclusive experts. Beyond the MoE-based method, there are alternative approaches like STAR \cite{star}, which introduces an extra tower for each scenario, combining the parameters of scenario-specific tower with those of the shared tower's parameter.

However, scenario-specific approaches learn scenario information transfer in quite implicit ways and often overlook differences in underlying representations, which frequently constitute a bottleneck in recommendation models. Therefore, SASS \cite{sass} designs a scenario adaptive transfer module to select and fuse effective transfer information from whole scenario to individual scenario. Adasparse \cite{adasparse} learns adaptively sparse structure for each scenario, achieving better generalization across domains with lower computational cost. HiNet \cite{hinet} achieves hierarchical extraction based on coarse-to-fine knowledge transfer scheme. DFFM \cite{dffm} incorporates scenario-related information into the parameters of feature interaction and user behavior modules, allowing scenario-specific learning of these two aspects, which is essentially a scenario-adaptive parameterization method. 3MN \cite{3mn} proposes a novel three meta networks-based solution to model the complicated task-task, scenario-scenario, and task-scenario interrelations. PEPNet \cite{pepnet} takes features with strong biases as input and dynamically scales the bottom-layer embeddings and the top-layer DNN hidden units in the model through a gate mechanism. MARIA \cite{maria} proposes to project the scenario semantic information into the bottom feature representation to derive more discriminative feature representations.

\section{Problem Formulation}
Given a set of scenarios $\mathcal{S}=\left\{s_m\right\}_{m=1}^{N_s}$ with a shared feature space $\mathcal{X}$ and label space $\mathcal{Y}$, the multi-scenario recommendation task aims to devise a unified ranking formula $\mathcal{F}: \mathcal{X} \rightarrow  \mathcal{Y}$, to concurrently provide accurate, personalized recommendations across $M$ scenarios. The common feature space typically encompasses a variety of features that can be categorized into distinct fields, such as user-centric and item-centric features, which collectively represent an instance's comprehensive contextual information. Within this paper, the feature space $\mathcal{X}$ is constructed as $[x_1, x_2, ..., x_{N_f}, \left\{b_i\right\}_{i=0}^{N_b}]$, where for clarity in subsequent discussions, $\left\{b_i\right\}_{i=0}^{N_b}$ denote the user's historical behavior sequence to distinguish it from other $N_f$ features, and $N_b$ represents the number of the user's historical behaviors. Mathematically, the multi-scenario recommendation task involves estimating the probability that the target user will interact with the target item in a given scenario $s_m$, utilizing the contextual features, as illustrated below:
\begin{gather}
\hat{y} = P(y=1| \mathcal{X}, s_m) =  \mathcal{F}(\boldsymbol{x}_1, \boldsymbol{x}_2, ..., \boldsymbol{x}_{N_f}, \left\{\boldsymbol{b}_i\right\}_{i=0}^{N_b}, \boldsymbol{s}_m)
\end{gather} 
where $\boldsymbol{x}_i \in \mathbb{R}^{d}$ denotes the embedding of the $i$-th feature, and $d$ denotes the embedding dimension of features. $\boldsymbol{b}_i \in \mathbb{R}^{d}$ denotes the representation of $i$-th behavior in the behavior sequence, typically derived from a pooled embedding that combines the clicked item embedding and the embeddings of its associated attribute features. $\boldsymbol{s}_m$ represents the representation vector of $m$-th scenario features. 

\begin{figure*}[t]
  \centering
  \includegraphics[width=\linewidth]{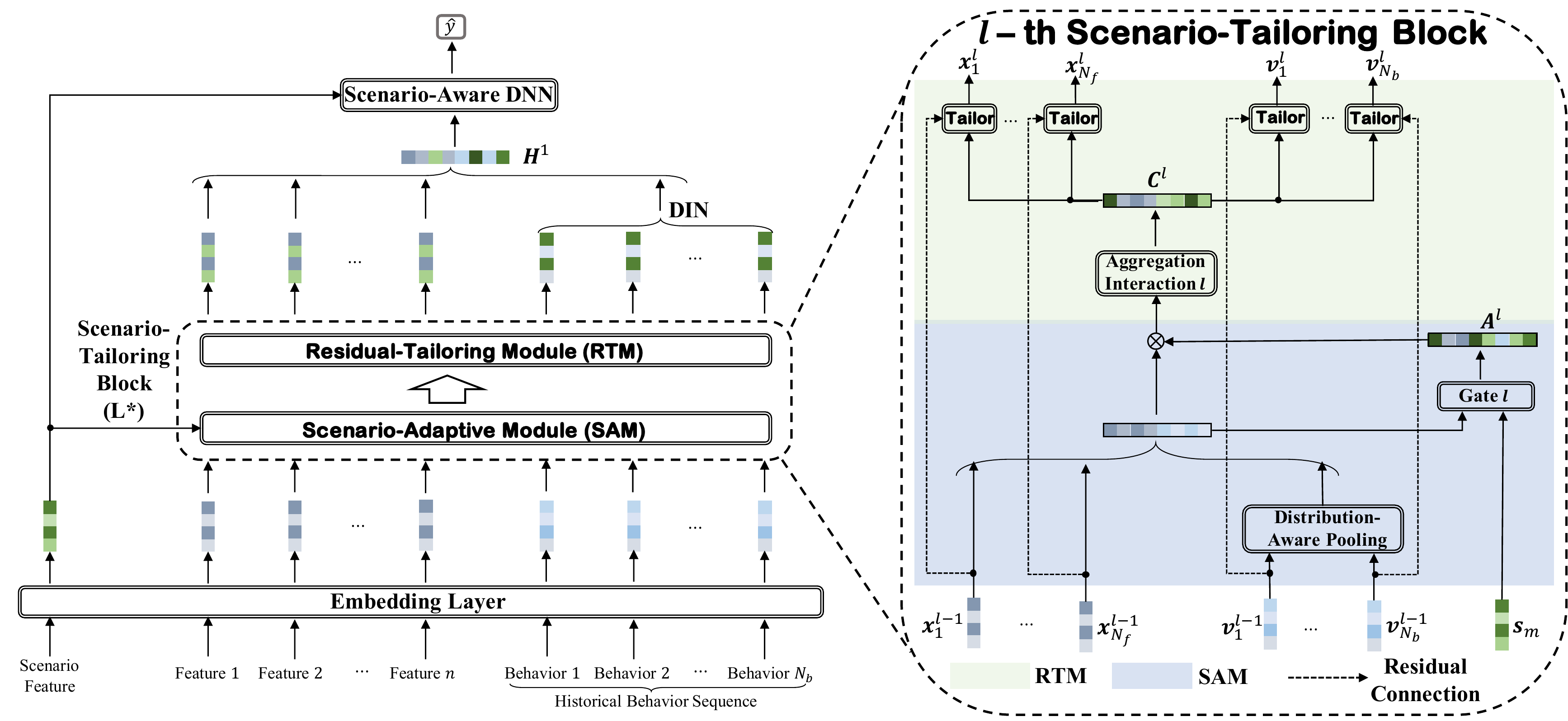}
  \caption{The network architecture of our proposed Scenario-Adaptive Fine-Grained Personalization Network (SFPNet).}
  \label{model}
\end{figure*}

\section{Method}
As previously discussed, contemporary multi-scenario models typically employ a unified framework to deliver personalized recommendations across various scenarios. While previous scene-adaptive parameter networks are lauded for their flexibility in adjusting the model's foundational representation to excel in multi-scenario recommendations, they typically consider the entire user historical behavior sequence as a singular feature field and conduct scenario-based transformations within this sequence. This approach uniformly adjusts all behaviors with the same weight, thus overlooking the distinct semantic information that each behavior may signify in different scenarios. In reality, users' interests frequently shift with the scenario transition, causing each behavior in the sequence to represent varying information across different scenarios. Inadequate cross-scenario modeling precludes precise extraction of user preferences, reflected in individual behaviors, for each scenario, thereby compromising the accuracy of cross-scenario recommendations. To more effectively model the intricate interplay between user historical behaviors and scenarios, this paper introduces a novel framework named Scenario-Adaptive Fine-Grained Personalization Network (SFPNet). It is primarily built from stacked \textbf{\emph{Scenario-Tailoring Block}} and comprises two main modules: \\
\textbf{$\bullet$ Scenario-Adaptive Module (SAM)}, which employs a gating mechanism to integrate scenario-related information into the base representations of features, ensuring that the instance's contextual information is strongly guided by the scenario context. \\
\textbf{$\bullet$ Residual-Tailoring Module (RTM)}, which aggregates scenario-aware contextual information and, via residual connections, disaggregates it into distinct representations for each behavior, allowing each behavior to capture contextual-cross information. 

\subsection{Scenario Adaptive Module (SAM)}  
Drawing inspiration from the LHUC algorithm \cite{lhuc}, which focuses on learning the distinct contributions of each speaker's hidden units, a variety of scenario-adaptive parameterization methods have been developed. These methods involve integrating different scenario prior bias features to create personalized network parameters for each instance, thereby enabling distinction of features at the foundational level across different scenarios. Consequently, within each Scenario-Tailoring Block, we also incorporate a Scenario-Adaptive Module (SAM) that coarsely applies scenario prior information to diverse feature representations. This ensures that while serving as contextual information for instances, the distinctions between scenarios are consistently maintained. Specifically, the input to the SAM in the first \emph{Scenario-Tailoring Block} is the original embedding vector of the features, and subsequent SAM inputs are typically the outputs from the \emph{Residual-Tailoring Module} of the previous block. As the purpose of SAM is to produce the contextual information for the instance, we compress the user behavior sequence into a distinct feature field, enabling complete perception of the sequence's information during subsequent behavior customization. SAM operates through a two-step process: sequence pooling, and gate personalization. To elucidate, let's examine the computation within SAM of the $l$-th block, delineated as follows:

\subsubsection{Sequence Pooling}  
Since that SAM's output serves solely as contextual information for the instance, our initial step is to pool the sequence. To avoid substantial increases in computational complexity from stacking blocks, we opt for computationally efficient aggregation methods—like summation, max pooling, or average pooling—for user sequences. This contrasts with the more complex and computationally intensive neural network structures used in DIN \cite{din} to capture sequence information related to target items. Nonetheless, the above simple pooling methods result in a measure of sequence information loss, meaning we are unable to accurately model the true distribution of sequences from the pooled sequence representations alone. For instance, considering the user sequence distribution adhering to a Gaussian distribution, we need to know both the mean and the variance to reconstruct the true sequence distribution, expressed as $b_i \verb|~| N(\mu, \sigma^2)$. Consequently, we refined the sequence pooling approach in this paper by developing a \emph{Distribution-Aware Pooling} method, which compresses the sequence while maximally preserving sequence information through the computation of both mean and variance, with minimal additional computational load. The precise computational procedure for pooling is delineated as follows:
\begin{gather}
\boldsymbol{B}^{l} =\left( \frac{1}{N_b} \sum_{i=1}^{N_b} \boldsymbol{b}_i^{l-1}, \sqrt{\frac{1}{N_b} \sum_{i=1}^{N_b} (\boldsymbol{b}_i^{l-1})^2 - (\frac{1}{N_b} \sum_{i=1}^{N_b} \boldsymbol{b}_i^{l-1})^2} \right ) 
\end{gather}
where $\boldsymbol{B}^{l} \in \mathbb{R}^{2d_{l-1}}$ denotes the pooled representation of user behavior sequence from the $(l-1)$-th block. $\boldsymbol{b}_i^{l-1} \in \mathbb{R}^{d_{l-1}}$ is the updated representation of $i$-th behavior output by the $(l-1)$-th block, and $d_{l-1}$ is the dimension of features in the $(l-1)$-th block.

\subsubsection{Gate Personalization}  
Given the scenario information for the $m$-th scenario, we have devised a gating function that applies scenario-specific adjustments at the instance level, tailored to each instance's features, to aid in the encoding of scenario semantics within instance feature representations. Specifically, we concatenate all features—encompassing instance-specific, compressed sequence, and scenario features as $\boldsymbol{X}^l=(\boldsymbol{x}_1^{l-1} \oplus \boldsymbol{x}_2^{l-1}\oplus \cdots \oplus \boldsymbol{x}_n^{l-1} \oplus \boldsymbol{B}^{l} \oplus \boldsymbol{s}_m) \in \mathbb{R}^{(n+2)d_{l-1}+d}$, where $\boldsymbol{x}_i^{l-1}$ denotes the representation of $i$-th feature from the $(l-1)$-th block—and employ two neural network layers to output a vector matching the dimensionality of the input instance features. The weight calculation for each element can be detailed as follows:
\begin{gather}
\boldsymbol{A}^l  =\sigma(\boldsymbol{W}_1^l(LN(ReLU(\boldsymbol{W}_0^l \boldsymbol{X}^l  + \boldsymbol{b}_0^l)))+ \boldsymbol{b}_1^l)
\end{gather}
where $\boldsymbol{A}^l \in \mathbb{R}^{(n+2)d_{l-1}}$ represents the weight of the element corresponding to the feature, ranging from $0$ to $1$. $\sigma(\cdot)$ denotes the sigmoid function. $LN(\cdot)$ denotes the layer-norm calculation, which aims to help the model learn a more differentiated weight distribution. $\boldsymbol{W}_0^l$, $\boldsymbol{W}_1^l$ and $\boldsymbol{b}_0^l$, $\boldsymbol{b}_1^l$ are the weight and bias for the gate network in the $l$-th block, respectively. 

Subsequently, we apply a re-weighting operation to the input representation of the $l$-th module within SAM, involving element-wise multiplication between the original representation and the calculated weight values, resulting in a new representation vector. Given that the gate's input encompasses scenario information, we can ensure that the re-weighted embeddings at each block layer consistently reflect scenario-specific distinctions. The specific computational formula is defined as follows:
\begin{gather}
\hat{\boldsymbol{X}}^l = (\boldsymbol{x}_1^{l-1} \oplus \boldsymbol{x}_2^{l-1}\oplus ... \oplus \boldsymbol{x}_n^{l-1} \oplus \boldsymbol{B}^{l}) \otimes \boldsymbol{A}^l
\end{gather}
In this way, $\hat{\boldsymbol{X}}^l \in \mathbb{R}^{(n+2)d_{l-1}}$ as an effective scenario signal will be further amplified in the following Residual-Tailoring Module, allowing the model to better capture scenario differences.

\subsection{Residual-Tailoring Module (RTM)}
As previously noted, previous multi-scenario modeling methods often differentiate features coarsely, thereby overlooking the intricate interplay of individual user behaviors with different scenarios. This limitation stymies the model's capacity to precisely track the evolution of user interests across scenarios, ultimately impairing the tailored recommendation of items that align with users' genuine interests in a cross-scenario context. To address the above issue, this paper introduces a Residual-Tailoring Module (RTM) that accepts the output of the Scenario-Adaptive Module (SAM). It leverages a neural network to facilitate interaction with context-aware features, producing a distinctive context representation vector for each feature. Additionally, residual connections \cite{res} are employed to integrate the contextual interaction vector into the representation vector of each user behavior, enabling fine-tuned scenario and context-aware customization for every behavior representation. RTM encodes the global semantics of user behavior within specific scenarios, thereby boosting the model's proficiency in detecting nuanced variations of user behavior across scenarios. It should be noted that while our module focuses on customizing user behavior, the same customization processes are applied to other features to augment their capacity to depict scene divergences. Specifically, the RTM comprises two principal steps: the aggregation interaction step and the context-aware tailoring step. The computation within $l$-th block of the RTM is delineated as follows:

\subsubsection{Aggregation Interaction Step}
Prior research \cite{cross} on feature interactions has demonstrated that deep neural networks adeptly capture implicit high-order feature interactions, frequently resulting in enhanced model prediction performance. Consequently, we integrate all scenario-aware contextual information from the input instance and utilize a neural network to dynamically produce high-order global feature interaction signals for each feature within the instance. This process unfolds as follows:
\begin{gather}
\boldsymbol{C}^l =\boldsymbol{W}_3^l (ReLU(\boldsymbol{W}_2^l \hat{\boldsymbol{X}}^l + \boldsymbol{b}_2^l)) + \boldsymbol{b}_3^l
\end{gather}
where $\boldsymbol{C}^l \in \mathbb{R}^{(n+2)d_{l}}$ denotes the contextual interaction information of the input features and include scenario information due to the input representations that have been modified to be adaptive to the scenario. $d_l$ is the dimension of the feature in $l$-th block. $\boldsymbol{W}_2^l$ and $\boldsymbol{W}_3^l$ are the weights of contextual information aggregation function, and $\boldsymbol{b}_2^l$ and $\boldsymbol{b}_3^l$ are the bias of function. 

\subsubsection{Context-Aware Tailor Step}
Drawing inspiration from residual architectures \cite{res}, we propose that once contextual interaction information is acquired via the aggregation module, residual connections are employed to merge the original feature representation with its corresponding global interaction representation, thus enabling recoding into a globally-aware representation. We begin by detailing the calculation of residual connections for all features, excluding sequences. The calculation process is as follows:
\begin{gather}
\boldsymbol{x}_i^{l} =ReLU(\boldsymbol{W}_{x_i, 1}^{l} \boldsymbol{x}_i^{l-1} + \boldsymbol{b}_{x_i, 1}^{l}) + \sigma(\boldsymbol{W}_{x_i, 2}^{l}\boldsymbol{x}_i^{l-1}  + \boldsymbol{b}_{x_i, 2}^{l}) \otimes \boldsymbol{c}_i^l
\end{gather}
where $\boldsymbol{x}_i^{l} \in \mathbb{R}^{d_{l}}$ denotes the context-aware scenario-adaptive representation of the $i$-th feature generated by the $l$-th block. $\boldsymbol{W}_{x_i, 1}^{l}$ and $\boldsymbol{W}_{x_i, 2}^{l}$ denote the weight matrix for the $i$-th feature. $\boldsymbol{x}_i^{l-1} = \boldsymbol{X}^{l-1}[i*d_{l-1}:(i+1)*d_{l-1}]$ denotes the scenario-adaptive representation of the $i$-th feature output by SAM. $\boldsymbol{c}_i^l = \boldsymbol{C}^l[i*d_l:(i+1)*d_l]$ denotes the context-aware cross information for the $i$-th feature. 

\textbf{Fine-Grained Cross-Scenario Sequence Modeling}. As emphasized throughout this paper, our goal with sequence features is to recode specific behavioral representations with fine granularity. The residual architecture is equally adaptable for fine-grained coding of sequence features. The specific computation involved in this fine-grained sequence coding is described as follows:
 \begin{gather}
\boldsymbol{v}_i^l =ReLU(\boldsymbol{W}_{4}^l \boldsymbol{v}_i^{l-1} + \boldsymbol{b}_{4}^l) + \sigma(\boldsymbol{W}_{5}^l \boldsymbol{v}_i^{l-1}  + \boldsymbol{b}_{5}^l) \otimes \boldsymbol{c}_b^l
\end{gather}
where $\boldsymbol{v}_i^{l} \in \mathbb{R}^{d_{l}}$ denotes the encoded representation of the $i$-th behavior of user sequence with awareness of context information and scenario information. $\boldsymbol{c}_b^l = \boldsymbol{C}^l[n*d_l:(n+2)*d_l]$ denotes the final context-aware scenario-adaptive representation of the sequence feature generated by the $l$-th block.  After re-encoding the sequence, each behavioral representation is finely tuned in accordance with the scenario. Aggregating the sequence at this juncture can substantially enhance the model's proficiency in extracting users' interests across various scenarios, thereby improving the model's adeptness at recognizing different situations. This, in turn, increases the accuracy of scenario-specific recommendations.

\begin{table*}[t]
	\small
	\centering
	\begin{tabular}{ccccccccccccc}
    		\hline Scenarios & \#A1 & \#A2 & \#A3 & \#A4 & \#A5 & \#A6 & \#A7 & \#A8 & \#A9 & \#A10 & \#A11 & \#A12  \\
		\hline \#User & 9,233,770 & 3,375,959 & 3,275,513 & 232,018& 3,613,010 & 4,066,414 & 4,474,204 & 2,968,181& 3,064,679 &783,071&1,352,973&157,645 \\
		\#Item & 1,309,072 & 1,150,596 & 1,138,230 & 1,038,012& 689,511 & 1,067,403 & 984,055 & 893,954&944,691 & 889,481&890,769&726,907 \\
		\#Click & 14,366,092 & 3,817,960 & 3,726,107 & 4,788,877& 5,156,626 & 2,279,444 & 1,498,577 & 1,662,265& 774,141 & 1,428,080 &1,869,587 & 490,689 \\
		% \#Impression & 613,063,353 & 184,225,328 & 166,313,438 & 156,479,014& 134,528,676 & 115,038,729 & 90,083,250 & 63,531,460 & 48,065,739 & 45,690,745 \\
		\#Impression & 613M & 184M & 166M & 156M& 135M & 115M & 90M & 64M & 48M & 46M & 44M & 29M \\
		\hline
	\end{tabular}
	\caption{Statistics of the industrial dataset.}
	\label{industrial_dataset}
\end{table*} 

\begin{table}[t]
	\small
	\centering
	\begin{tabular}{ccccc}
    		\hline Scenarios & \#B1 & \#B2 & \#B3 & \#All \\
		\hline \#User & 91,488 & 2,612 & 154,024 & 244,397  \\
		\#Item & 535,711 & 198,651 &537,937 & 538,376 \\
		\#Click & 1,291,063 & 28,022 &1,998,618 & 3,317,703 \\
		\#Impression & 32,236,951 & 639,897 &52,439,671 & 85,316,519 \\
		\hline
	\end{tabular}
	\caption{Statistics of the Ali-CCP dataset.}
	\label{public_dataset}
\end{table}

\subsection{Scenario-Specific Prediction}  
In the prediction layer, the sequence must first be compressed to facilitate its integration with other feature vectors for subsequent processing by the deep network. Given the article's focus on sequence representation recoding, we employ commonly used structures for the compression vector, such as the target-attention mechanism of Deep Interest Network (DIN) \cite{din}:
\begin{gather}
\boldsymbol{V} =DIN(\boldsymbol{v}_1^L, \boldsymbol{v}_2^L, ..., \boldsymbol{v}_{N_b}^L)
\end{gather}
where $L$ is the total number of block, and $\boldsymbol{v}_i^L$ is the representation of $i$-th behavior outputted by the last block. Subsequently, we leverage the benefits of a multi-tower architecture by incorporating scenario-aware DNN towers at the prediction stage. These towers process sequence features and other features in parallel by considering the scenario bias each layer, to forecast the likelihood of users clicking on the target item within the specific scenario $s_m$:
\begin{gather}
\hat{y} =P(y=1| \mathcal{X}, s_m) = \sigma(\mbox{\emph{S-DNN}}(\boldsymbol{H}^1, \boldsymbol{s}_m)) 
\end{gather}
where $\boldsymbol{H}^1 = (\boldsymbol{x}_1^L \oplus \boldsymbol{x}_2^L \oplus \cdots \oplus \boldsymbol{x}_n^L  \oplus \boldsymbol{V})$ denotes the input of the first layer of the Scenario-Aware DNN (S-DNN). The specific calculation of the $j$-th layer in the S-DNN structure is similar to the calculation in the SAM, and can be depicted as follows:
\begin{gather}
\boldsymbol{A}^j_p =\sigma(\boldsymbol{W}_7^j(LN(ReLU(\boldsymbol{W}_6^j (\boldsymbol{H}^{j-1} \oplus \boldsymbol{s}_m)  + \boldsymbol{b}_6^j)))+ \boldsymbol{b}_7^j) \\
\boldsymbol{H}^{j} = ReLU(\boldsymbol{W}_8^j (\boldsymbol{H}^{j-1} \otimes \boldsymbol{A}^j_p )+ \boldsymbol{b}_8^j)
\end{gather}
where $\boldsymbol{H}^{j}$ denotes the output of the $j$-th layer of S-DNN. $\boldsymbol{W}_6^j$, $\boldsymbol{W}_7^j$ and $\boldsymbol{W}_8^j$ are the weight for the S-DNN in the $j$-th layer. $\boldsymbol{b}_6^j$, $\boldsymbol{b}_7^j$ and $\boldsymbol{b}_8^j$ are the bias for S-DNN in the $j$-th layer. 

\subsection{Model Learning}  
The objective function applied in our model is the cross entropy loss function, defined as:
\begin{gather}
\mathcal{L} = -\frac{1}{N} \sum_{i=1}^N (y_i log(\hat{y}_i) + (1-y_i)log(1-\hat{y}_i))
\end{gather} 
where $y_i \in \left\{0, 1\right\}$ is the ground truth of instance.

\section{Experiments}
In this section, we conduct extensive experiments on public bench- mark datasets to validate the effectiveness of our proposed framework and answer the following questions: \\
\textbf{$\bullet$ RQ1:} How does SFPNet perform compared with the state-of-the-art baseline methods?  \\
\textbf{$\bullet$ RQ2:} How does the performance of the SFPNet model, trained on specific scenario data for each scene, compare to that of the SFPNet model trained on a composite scenario dataset? \\
\textbf{$\bullet$ RQ3:} How about the impact of each part on the overall model? \\
\textbf{$\bullet$ RQ4:} How about the impact of hyper-parameters of our model?

\subsection{Datasets}
To validate the efficacy of our proposed method, two real-world large-scale datasets covering diverse scenarios are used for performance evaluation. We conduct experiments on both public available dataset and industrial dataset. The descriptions and statistics of two dataset are detailed in Table \ref{industrial_dataset} and \ref{public_dataset}, respectively. \\
\textbf{$\bullet$ Industrial Dataset}. To assess our multi-scenario recommendation approach in a real-world setting, we gathered a dataset from the Alibaba international advertising platform, Lazada, that spans multiple scenarios from December 10, 2023, to January 10, 2024. This dataset encompasses 12 scenarios, denoted as \#A1 through \#A12. We allocated the user instances from the final day for testing and the preceding instances for training. \\
\textbf{$\bullet$ Ali-CCP  \footnote{https://tianchi.aliyun.com/dataset/408}}.  It is widely used in the relevant literature for multi-scenario recommendations \cite{maria, plate, sass}, which was collected Taobao’s recommender system under three scenarios. It's released by Taobao with prepared training and testing set, and we can split the dataset into 3 scenarios according to the \emph{scenario id}, denoted as \#B1, \#B2, and \#B3 for simplicity.

\subsection{Competitors}
We conduct experiments with several compared methods for the multi-scenario recommendation task. These methods can be divided into three groups as follows:
\subsubsection{General Recommenders} Samples from all scenarios are utilized to co-train a universal recommender system capable of multi-domain recommendations. \\
\textbf{$\bullet$ BaseDNN}.  It is a multi-scenario model that shares the parameters of the bottom layer. On top of the shared bottom layer, a single DNN is used for prediction across scenarios. \\
\textbf{$\bullet$ DeepFM} \cite{deepfm}.  The model resembles BaseDNN; however, it substitutes the DNN architecture with DeepFM, which is a neural network framework grounded in factorization machines. \\
\textbf{$\bullet$ SharedBottom}.  It replaces the single DNN with multiple DNNs. That is, an individual DNN is utilized for each scenario. And we add an auxiliary tower to enhance the ability to characterize the scenario indicator. 

\subsubsection{Scenario-specific network structures} Each scenario is treated as a distinct task and inter-scenario correlations are examined through dedicated networks for each scenario. \\
\textbf{$\bullet$ MMoE} \cite{mtl3}.  It  implicitly models task relationships for multi-task learn- ing, where different tasks may have different label spaces. Here we adapt MMoE for multi-scenario learning. The number of experts is equal to the number of experts of Maria. The sum of weighted outputs from the experts are fed into the individual tower for each scenario respectively. \\
\textbf{$\bullet$ PLE} \cite{ple}.  It is a state-of-the-art multi-scenario/multi-task model that organizes the experts into scenario-specific groups and scenario- shared groups for the purpose of avoiding negative transfer or seesaw phenomenon. \\
\textbf{$\bullet$ STAR} \cite{star}.  It proposes a star topology to accommodate with the scenario-specific characteristics. Specifically, a shared network works as the center node for knowledge sharing and each scenario network connects only with the center node. \\
\textbf{$\bullet$ AESM$^2$}  \cite{aesm2}.  It proposes a novel MMoE-based model with automatic search towards the optimal network structure. In contrast to PLE and STAR, an expert can be either scenario-shared or scenario- specific dynamically in an instance-aware manner.

\subsubsection{Scenario-adaptive parameter network structures} They disregard feature space differences and applying scenario information directly to the core embedding layer and prediction module, thus allowing the feature space and prediction strategies to dynamically adjust to scenario variations.\\
\textbf{$\bullet$ AdaSparse} \cite{adasparse}.  It learns adaptively sparse structures for multi-scenario prediction and prunes redundant neurons via learned neuron-level weighting factors to improve generalization. \\
\textbf{$\bullet$ PEPNet} \cite{pepnet}.  It takes features with strong biases as input and dynamically scales the bottom-layer embeddings and the top-layer DNN hidden units in the model through a gate mechanism. \\
\textbf{$\bullet$ MARIA} \cite{maria}. It designs three components to enable discriminative feature learning in a scenario-aware manner: feature scaling, feature refinement, and feature correlation modeling. 

\begin{table*}[t]
	\caption{Prediction performance on the industrial dataset. * indicates p-value < 0.05 in the significance test.}
	\begin{tabular}{c|c|c|c|c|c|c|c|c|c|c|c|c}
    \toprule
    \multirow{3}{*}{Method}&
    \multicolumn{12}{c}{The industrial dataset (S-GAUC)}\cr
   %  \cmidrule(lr){2-7} \cmidrule(lr){8-13}
    \cmidrule(lr){2-13}
    &\#A1&\#A2&\#A3&\#A4&\#A5&\#A6&\#A7&\#A8&\#A9&\#A10&\#A11&\#A12\cr
    \midrule
    BaseDNN & 0.6396 & 0.5501 & 0.6313  & 0.6322 & 0.6514 & 0.6176   & 0.6494 & 0.6571 & 0.6471 & 0.6390 & 0.6212 & 0.6298 \cr
    DeepFM  & 0.6398 & 0.5517 & 0.6309  & 0.6318 & 0.6523 & 0.6181   & 0.6499 & 0.6573 & 0.6468 & 0.6388 & 0.6208 & 0.6293  \cr
    SharedBottom  & 0.6352 & 0.5496 & 0.6313 & 0.6294 & 0.6486  & 0.6115  & 0.6437 & 0.6526 & 0.6415 & 0.6339& 0.6192 & 0.6273 \cr
    \midrule
    MMoE    & 0.6424 & 0.5508 & 0.6321  & 0.6356 & 0.6558 & 0.6157 &  0.6535 & 0.6592 & 0.6499 & 0.6425& 0.6234 & 0.6308 \cr
    PLE   & 0.6456 & 0.5529 & 0.6348    & 0.6353  & 0.6581  & 0.6175 & 0.6551 & 0.6602 & 0.6513 & 0.6457& 0.6268 & 0.6317    \cr
    STAR   &0.6412  & 0.5512 & 0.6304     & 0.6349  & 0.6536  & 0.6159 & 0.6529 & 0.6579 & 0.6481 & 0.6413 & 0.6217 & 0.6279   \cr
    AESM$^2$  &0.6448  &0.5524  & 0.6343  &0.6341  & 0.6567  & 0.6168 & 0.6538 & 0.6579 & 0.6501 & 0.6423  & 0.6251 & 0.6301  \cr
    \midrule
    AdaSparse   & 0.6450 & 0.5531 & 0.6374  & 0.6362  & 0.6582  & 0.6201 & 0.6573 & 0.6605 & 0.6505  & 0.6451& 0.6282 & 0.6313 \cr
    PEPNet & 0.6463 & 0.5563 & 0.6402 & 0.6385  & 0.6595 & 0.6235  & 0.6612& 0.6636 & 0.6527 &0.6492 & 0.6306 & 0.6350   \cr 
    MARIA  & 0.6472 & 0.5561 & 0.6418  & 0.6392  & 0.6601 & 0.6242 &0.6615 & 0.6647 & 0.6531 & 0.6495& 0.6312 & 0.6358  \cr
    \midrule
    SFPNet & \textbf{0.6524}* & \textbf{0.5612}* & \textbf{0.6489}* &  \textbf{0.6447}* & \textbf{0.6658}* & \textbf{0.6309}* & \textbf{0.6689}* & \textbf{0.6714}* & \textbf{0.6597}* & \textbf{0.6548}*& \textbf{0.6375}* & \textbf{0.6432}* \cr
    \bottomrule
    \end{tabular}
	\label{overall}
	\label{result_ind}
\end{table*} 

\begin{table}[t]
	\caption{Prediction performance on the Ali-CCP dataset. * indicates p-value < 0.05 in the significance test.}
	\begin{tabular}{c|c|c|c}
    \toprule
    \multirow{3}{*}{Method}&
    \multicolumn{3}{c}{Ali-CCP dataset (AUC)}\cr
   %  \cmidrule(lr){2-7} \cmidrule(lr){8-13}
    \cmidrule(lr){2-4}
    &\#B1&\#B2&\#B3\cr
    \midrule
    BaseDNN & 0.5723 & 0.5887 & 0.5869 \cr
    DeepFM  & 0.5741 & 0.5903 & 0.5878  \cr
    SharedBottom  & 0.5685 & 0.5881 & 0.5817 \cr
    \midrule
    MMoE    & 0.5764 & 0.5916 & 0.5884 \cr
    PLE   & 0.5786 & 0.5828 & 0.5893   \cr
    STAR   & 0.5672 & 0.5869 & 0.5803   \cr
    AESM$^2$   & 0.5798 & 0.5963 & 0.5927   \cr
    \midrule
    AdaSparse   & 0.5808 & 0.5986 & 0.6024\cr
    PEPNet & 0.5839 & 0.6101 & 0.6052  \cr 
    MARIA  & 0.5864 & 0.6108 & 0.6068  \cr
    \midrule
    SFPNet & \textbf{0.5917}* & \textbf{0.6153}* & \textbf{0.6105}*  \cr
    \bottomrule
    \end{tabular}
	\label{overall}
	\label{result_pub}
\end{table} 

\textbf{Implementation Details}. In offline experiments, we implement all the models based on the TensorFlow framework \cite{tensorflow}. We use Adam \cite{adam} for optimization with an initial learning rate of 0.001 and a decay rate of 0.9. The batch size is set as 512 and the embedding size is fixed to 40 for all models. Xavier initialization \cite{xavier} is used here to initialize the parameters. All methods use a three-layer feedforward neural network with hidden sizes of [256, 128, 64] for instance prediction. We apply careful grid-search to find the best hyper-parameters. The number of experts in MMoE, PLE, SharedBottom and its variants is searched in [2, 4, 6, 8]. All regularization coefficients are searched in [1e-7, 1e-5, 1e-3].

\textbf{Evaluation Metric}. In line with prior studies, this paper employs the area under the ROC curve (AUC) as the performance metric for the public Ali-CCP dataset. However, for our industrial dataset, we adopt a variant of session-weighted AUC. This metric assesses the quality of item rankings within sessions by averaging the AUCs of a user's individual session behaviors. This variant, already implemented on our platform, has proven to align more closely with the system's online performance. For brevity, we refer to this metric as S-GAUC, and its computation is as follows:
\begin{gather}
\mbox{\emph{S-GAUC}} = \frac{\sum_{i=1}^n \#impression_i \times AUC_i} {\sum_{i=1}^n \#impression_i}
\end{gather}
where $n$ is the number of sessions in the dataset. $\#impression$ and $AUC_i$ are the number of impressions and AUC corresponding to the $i$-th session. A higher $\mbox{\emph{S-GAUC}}$ score usually denotes a better online recommendation performance. We conduct a Mann-Whitney U test \cite{m-test} under AUC and S-GAUC metrics. 

\subsection{Comparison with Baselines (RQ1)}
Table \ref{result_ind} and Table \ref{result_pub} displays the overall prediction performance of all methods on the industrial and public dataset, respectively, along with the statistical significance of our model against the best baseline model, with the highest results highlighted in bold. From the results in two tables, we can see that SFPNet outperforms all baselines for both datasets, indicating that the Scenario-Tailoring Block can lead to multi-scenario prediction performance improvements.

As for \emph{General Recommenders}, we can find that they often struggle to attain satisfactory performance compared to other baseline models on the both datasets. SharedBottom incorporates dedicated DNNs for each scenario to embed scene-specific biases into the model. However, empirical findings indicate that this straightforward approach of directly incorporating biases can frequently yield detrimental outcomes. This outcome corroborates our hypothesis that bottom-level sharing with a single top structure is inadequate for capturing the complex interplay between scenarios. Consequently, with datasets that exhibit significant scene variability, the performance of \emph{General Recommenders} tends to be subpar.  

To address the shortcomings of \emph{General Recommenders}, \emph{Scenario-Specific Network Structure Methods} were proposed. According to experimental outcomes, these methods have resulted in notable performance enhancements. By enhancing the model's capacity to distill shared knowledge across scenarios, both the MMoE and STAR models outperformed the previously mentioned three baseline models. However, similar to the seesaw effect observed in multi-task learning, MMoE and PLE also exhibit a seesaw effect across multiple scenarios. This indicates that these structures alone are insufficient for efficiently handling scenarios with uneven data distribution, and achieving simultaneous performance improvements across all scenarios. To mitigate the seesaw effect, PLE, a variant of MMoE, introduces greater stability across scenarios by segmenting the expert network into two distinct groups. Notably, PLE demonstrates significant improvement over MMoE on the Ali-CCP dataset. Additionally, AESM$^2$ performs well on the Ali-CCP dataset, yet it underperforms on the industrial dataset. We attribute this to the extensive variety of industrial scenarios. AESM$^2$ mechanism for automatically selecting expert networks faces challenges in training with such a diversity of scenes. This suggests that in more complex scenarios, the comparatively simpler strategy employed by PLE may more readily achieve optimal performance.

Despite the focus on optimization at the top level by the aforementioned methods, some studies have indicated that the bottom level often represents a bottleneck in recommendation systems \cite{maria, pepnet}.  As evidenced by the results in Tables \ref{result_ind} and \ref{result_pub}, the \emph{Scenario-Adaptive Parameter Methods} delivered commendable predictive performance on both datasets, lending support to this hypothesis. Notably, Adasparse's performance is marginally inferior to that of PEPNet. We surmise that this is because the mechanism of sparse neurons is relatively challenging to learn. Sparse data scenarios in recommendations frequently result in suboptimal performance. The MARIA model outshines others on both datasets by optimizing both the lower and upper structures of the model. However, it is apparent that these methods do not fully address the pivotal task of modeling user interests across multi-scenarios. This oversight is why our approach, SFPNet, secures the best prediction performance across both datasets, emphatically underscoring the significance of finely-grained modeling of user behaviors across scenarios.

\subsection{Comparison with Training Alone (RQ2)}
\begin{figure}[t]
  \centering
  \includegraphics[width=\linewidth]{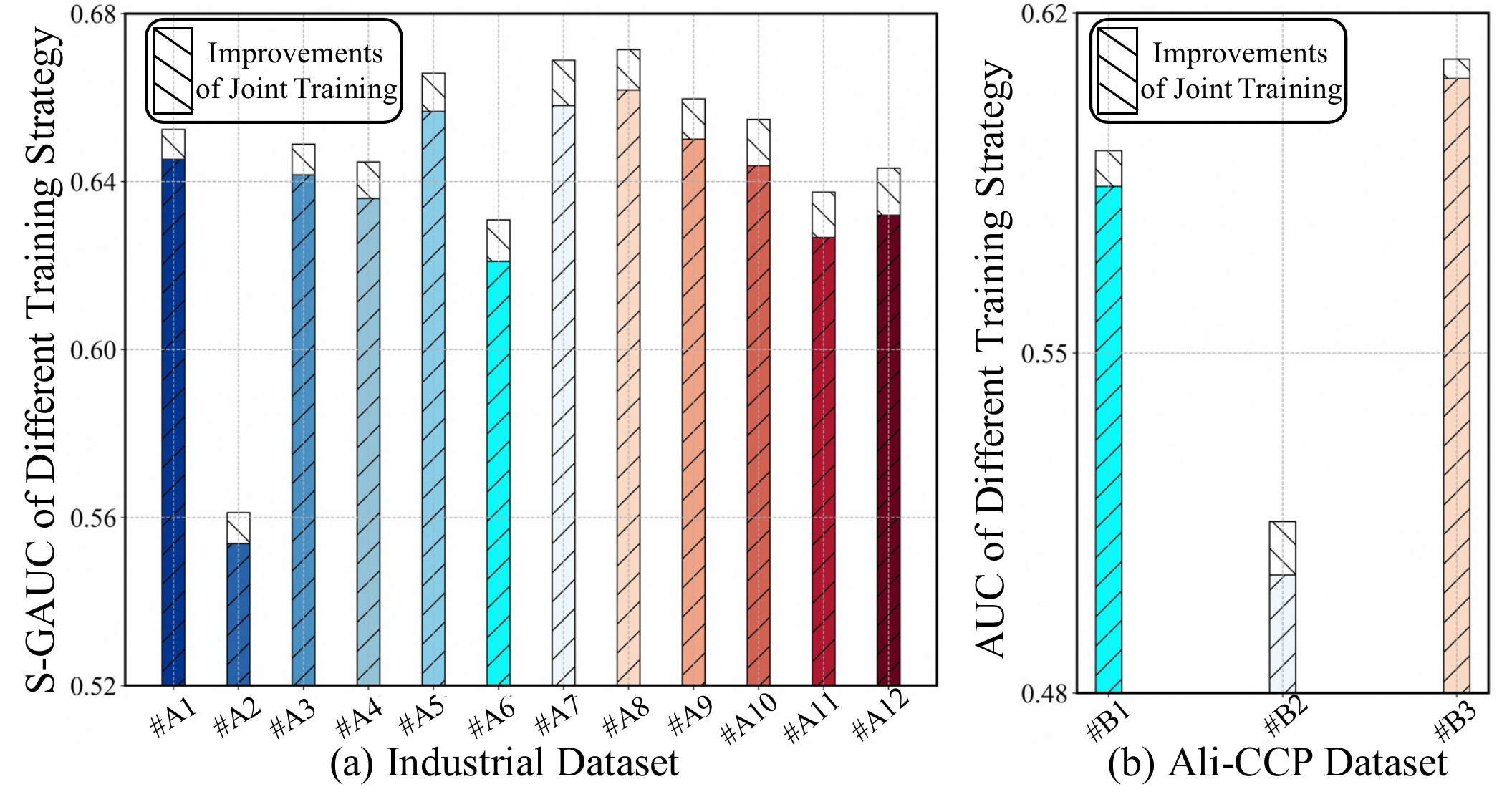}
  \caption{The impact of multi-scenario joint training.}
  \label{data_sep}
\end{figure}

While our model SFPNet surpasses the baselines in scenario-based prediction accuracy through multi-scenario joint modeling, we must ascertain whether multi-scenario joint training confers benefits on each individual scenario. Specifically, we question whether training a separate model for each scenario using our custom module might yield better results than joint training. Should some scenarios experience diminished performance following joint training, it may be advantageous to model these scenarios independently to achieve better online outcomes. This would also indicate that our model has room for improvement in accurately modeling inter-scenario relationships. Consequently, to investigate the potential issues outlined above, we trained SFPNet individually on both public and industrial datasets. The detailed experimental results, illustrated in the Figure \ref{data_sep}, show a white box representing improved performance when employing joint modeling as compared to separate training. The results reveal that independent training not only requires more manpower and resources but also underperforms compared to joint training. These findings underscore the importance of leveraging data from various scenarios for recommendation tasks, particularly those with sparse data, and also validate our multi-scenario structural design's ability to effectively transfer knowledge between scenarios.

\begin{table*}[t]
	\caption{The performance of five variants on the industrial dataset. We record the mean results over 5 runs. }
	\begin{tabular}{c|c|c|c|c|c|c|c|c|c|c|c|c}
    \toprule
    \multirow{3}{*}{Method}&
    \multicolumn{12}{c}{The industrial dataset (S-GAUC)}\cr
   %  \cmidrule(lr){2-7} \cmidrule(lr){8-13}
    \cmidrule(lr){2-13}
    &\#A1&\#A2&\#A3&\#A4&\#A5&\#A6&\#A7&\#A8&\#A9&\#A10&\#A11&\#A12\cr
    \midrule
    SFPNet & \textbf{0.6524} & \textbf{0.5612} & \textbf{0.6489} &  \textbf{0.6447} & \textbf{0.6658} & \textbf{0.6309} & \textbf{0.6689} & \textbf{0.6714} & \textbf{0.6597} & \textbf{0.6548}& \textbf{0.6375} & \textbf{0.6432} \cr
    w/o DAP & 0.6506 & 0.5596 & 0.6478 & 0.6429 & 0.6649 & 0.6301 & 0.6677 & 0.6701 & 0.6583 & 0.6531 & 0.6364 & 0.6413 \cr
    w/o SAM  & 0.6491 & 0.5573 & 0.6462 & 0.6403 & 0.6621 & 0.6284 & 0.6651 & 0.6673 & 0.6558 & 0.6512 & 0.6339 & 0.6401  \cr
    w/o RTM  & 0.6470 & 0.5560 & 0.6424 & 0.6397 & 0.6605 & 0.6250 & 0.6621 & 0.6655 & 0.6528 & 0.6501  & 0.6318 & 0.6364    \cr
    w/o S-DNN & 0.6497 & 0.5582 & 0.6470 & 0.6414 & 0.6628 & 0.6291 & 0.6660 & 0.6678 & 0.6567 & 0.6526 & 0.6351 & 0.6426  \cr
    w/o STB  & 0.6461 & 0.5553 & 0.6415 & 0.6384 & 0.6596 & 0.6241 & 0.6613 & 0.6643 & 0.6515 & 0.6492  & 0.6301 & 0.6350    \cr
    \bottomrule
    \end{tabular}
	\label{overall}
	\label{ablation}
\end{table*} 

\subsection{Ablation Study (RQ3)}
To verify the effectiveness of each module in the proposed model, i.e., the SFPNet method, we conduct a series of ablation studies over the industrial dataset. We have five variants as follows: \\ 
\textbf{$\bullet$ w/o DAP}  removes the distribution-aware sequence pooling method and directly adopts the average pooling of sequences. \\
\textbf{$\bullet$ w/o SAM} removes the Scenario-Adaptive Module, which means that the block no longer emphasizes scenario information. \\
\textbf{$\bullet$ w/o RTM}  removes the Residual-Tailoring Module in each block, which implies that the block no longer provides finely-tuned, scenario-specific representations for individual user behaviors. \\
\textbf{$\bullet$ w/o S-DNN} removes the scenario-aware DNN and replace it with a normal deep neural network. \\
\textbf{$\bullet$ w/o STB} removes the Scenario-Tailoring Block and directly conducts predictions based on S-DNN . 

Table \ref{ablation} presents the performance metrics of five SFPNet variant models. The results indicate that SFPNet's performance diminishes with the removal of any module, underscoring the contribution of each component to predictive accuracy. Specifically, the absence of distribution-aware sequence modeling (w/o DAP) impacts prediction performance across all scenarios, aligning with our hypothesis that mean pooling in sequence modeling may result in a loss of critical information. Additionally, we observed that despite the presence of an S-DNN atop SFPNet to assist with scene perception, the removal of SAM (w/o SAM) still results in a decline in model performance, reinforcing our understanding of the critical nature of the bottom representation proposed in previous work—akin to the proverb "a breach in a dike may cause a collapse". Furthermore, the Residual-Tailoring Module's absence (w/o RTM) notably affects the performance of each scenario. Indeed, without the RTM, our model reverts to the scene adaptive parameter structure of prior multi-scenario modeling methods, consequently forfeiting the capability for scenario-specific customization of user behavior. The experimental data suggests that the performance drop from removing either the RTM or the Scenario-Tailoring Block (w/o STB) is quite comparable. This underscores the significance of the RTM and robustly validates our rationale for concentrating on fine-grained scenario modeling exploration. Lastly, the removal of S-DNN (w/o S-DNN) also leads to a measurable reduction in the model's prediction performance. This reveals that, although we prioritize the bottom layer's architecture, the top structure's design must not be overlooked. This reflects the initial design motivation for the method, which was to clearly structure the scenario.
 
\begin{figure}[t]
  \centering
  \includegraphics[width=\linewidth]{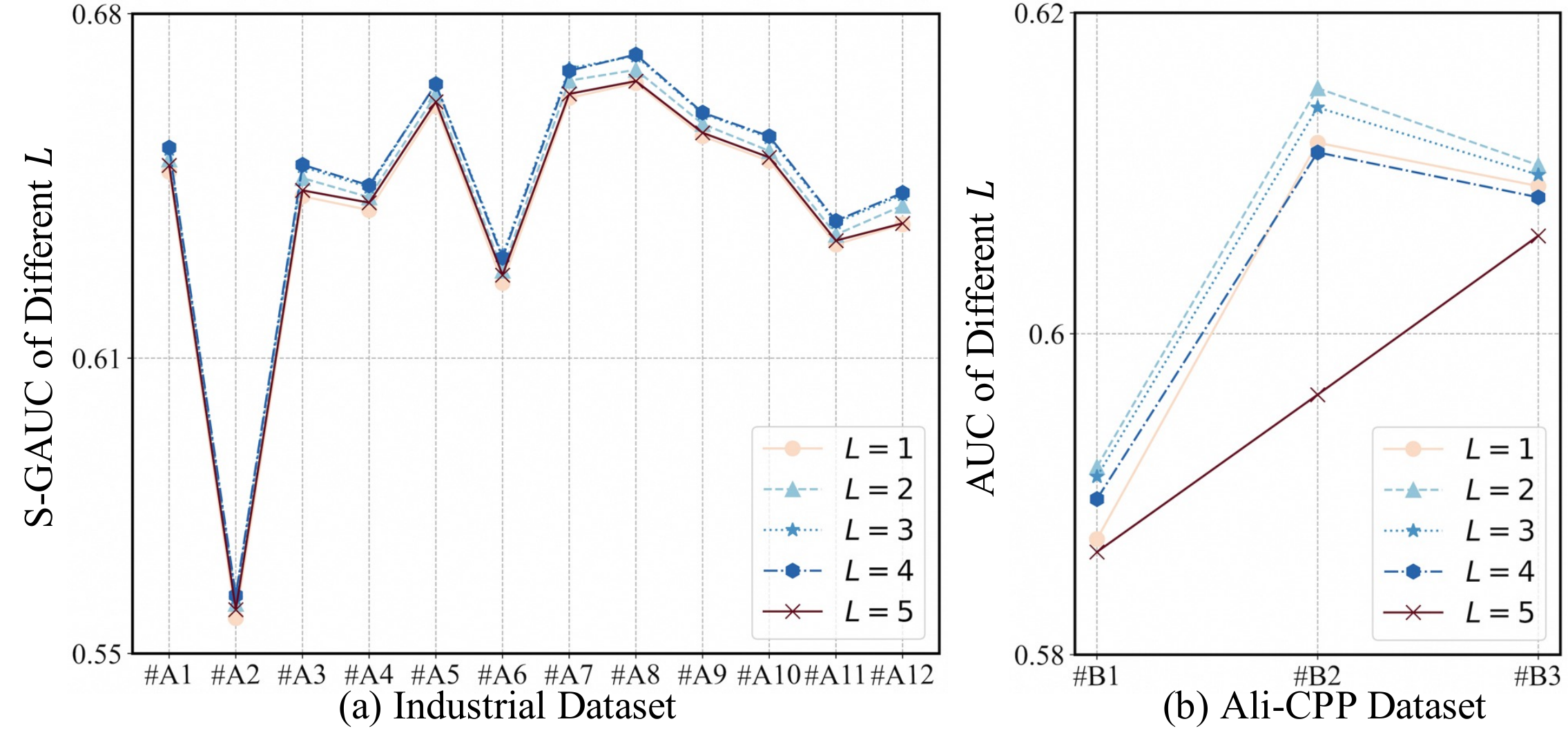}
  \caption{The performance of SFPNet under different predefined hyper-parameters on the both datasets.}
  \label{hyper}
\end{figure} 

\subsection{Hyper-Parameters Sensitivity Analysis (Q4)}
This section examines the sensitivity of the hyper-parameter of Scenario-Tailoring Block stacking times within the SFPNet model, i.e., $L$. We assessed SFPNet's performance on two datasets with five different values of $L$ (1, 2, 3, 4, and 5), with results presented in Figure \ref{hyper}(a) for the industrial dataset and Figure \ref{hyper}(b) for the other. Analysis of the industrial dataset reveals that $L$ values of 3 and 4 yield similar and optimal prediction performance; however, an additional block layer increases computational demands. Consequently, we recommend setting $L$ to 3. Theoretically, increasing the number of stacks should deepen the model, potentially enhancing its capability to model multiple scenarios. Nevertheless, given the dataset size, elevating the stack count beyond a certain point complicates training and can deteriorate prediction performance. Similarly, for the public dataset, setting $L=2$ results in peak performance for SFPNet. This optimal setting is attributable to the smaller size of the Ali-CCP dataset compared to the industrial dataset. Excessive block stacking increases the likelihood of overfitting in the model.

\subsection{Online A/B Testing Results}
To more robustly validate our model's performance, we conducted an online A/B test on an online e-commerce platform. The control in our test was a base model—a model structure corresponds to the BaseDNN from our baseline comparisons. We averaged the results of the 12 scenarios in our platform from October 21 to 26, 2023, to determine the final outcome of the online test. Specifically, following the implementation of the SFPNet model, we observed a \emph{\textbf{6.4\%}} increase in cumulative \emph{\textbf{Revenue}} and a \emph{\textbf{9.2\%}} rise in user \emph{\textbf{Click-Through Rate (CTR)}} compared to the base model. The online results further corroborate the effectiveness of the proposed SFPNet approach for multi-scenario recommendation. 

\section{Conclusions}
In this paper, we point out the optimizable space for modeling user sequences within multi-scenario contexts in the multi-scenario recommendation field and propose a novel model named Scenario-Adaptive Fine-Grained Personalization Network (SFPNet) to explore the optimization space for modeling user sequences with awareness of scenarios in a fine-grained way. SFPNet is is composed of the Scenario-Tailoring Block stacked block by block, and each block comprises two key components: the Scenario-Adaptive Module (SAM) and the Residual-Tailoring Module (RTM). Specifically, SAM employs a gating mechanism to infuse scenario-specific information into the instance's contextual features, whereas RTM harnesses the generated scenario-aware contextual information by SAM to craft a distinctive representation vector for each behavior in the sequence, utilizing residual connections for this purpose. This process ultimately bolsters the model's capacity to capture the evolving patterns of user interests across different scenarios. Finally, extensive experiments validate  the superiority of SFPNet in the multi-scenario recommendation task.

\end{document}